# The secondary flow in a short aspect ratio circular lid driven cavity at small but finite Reynolds number

**Benson K. Muite**[*]

School of Engineering and Applied Science
Princeton University
Princeton, NJ08544, USA

*Regular perturbation solutions are obtained for the Stokes flow field, the first order effects of inertia on the flow field, and the primary and secondary pressure fields in the circular lid driven cavity. The physical mechanism that causes vortex breakdown exists at small Reynolds number; it is a stagnation of the secondary flow by an adverse pressure gradient. The discontinuity between the rotating lid and the stationary sidewall has negligible influence on the flow field provided that the inertial forcing of the secondary flow is not localized near the boundary discontinuity, and the Reynolds number is small.*

## I. Introduction

The circular lid driven cavity (see Fig. 1) is a well-studied flow because it has a Reynolds number and aspect ratio, $\gamma = L^*/r_o^*$ (where $L^*$ is the length of the cavity and $r_o^*$ its radius) dependent vortex breakdown[1,2]. Two geometries have been studied, one with a stationary solid surface opposite the rotating endwall[1,2], and the other with a free surface opposite the rotating endwall[2]. For Reynolds numbers less than 500, the dominant velocity component is azimuthal. There is also a secoondary axial and radial flow in which: fluid leaves the upper rotating lid, travels down the stationary sidewall, moves in along the bottom wall, and finally returns up along the center of the cavity. When the Reynolds number is greater than a thousand with a fixed upper surface, or greater than 500 with a free surface, and the aspect ratio between one and two, experiments show the eddy pattern that exists at lower Reynolds numbers is modified[1,2]. Along the central axis of the cavity, the secondary flow has a stagnation point. After the stagnation point follows a bubble like region with circulation in the axial and radial plane that is in the opposite direction to circulation in the rest of the cavity.

This work was stimulated by an experimental investigation of the effect of secondary forcing on the flow field in a circular lid driven cavity by Sinha[3]. In Sinhas' investigation, the rotating endwall undulated over a less than 5° angle. The extra vorticity created by this undulating endwall changed the structure of the bubble in the center of the cavity. At present, there are questions on the exact structure of the bubble because there is poor agreement between experiment and simulation[2].

---

[*] Corresponding author. Telephone: +254-0721-732-876; fax: +254-20-572061;
electronic mail: benson_muite@yahoo.com
Present address: PO Box 67920-00200, Nairobi, Kenya

There have been several previous investigations of the Stokes flow in a cylindrical lid driven cavity with a fixed endwall opposite the rotating endwall. Panton[4], has found a perturbation solution for the Stokes flow field in a short aspect ratio geometry. Khalili and Rath[5] have also investigated exact Stokes flow solutions for the cylindrical lid driven cavity. Pao[6] has considered the time dependent Stokes flow of a rotating container of fluid, the bottom of which is suddenly brought to rest. Pao[6] experimentally and numerically verified that the Stokes solution described the azimuthal flow for Reynolds numbers less than ten. Pao[6] was aware that the discontinuity may affect his calculations and on examining his numerical results, noted that nothing peculiar occurred at the boundary discontinuity for the range of Reynolds numbers he studied.

Azerad and Bänsch[11] studied the flow in a cone and plate rheometer. This flow is similar to the circular lid driven cavity considered here, except that instead of a top rotating lid, there is a rotating cone with a tip that extends to the stationary base of the cavity. This geometry also has a boundary discontinuity that Azerad and Bänsch[11] rigorously show can be neglected at low Reynolds number. Azerad and Bänsch[11] provide an abstract proof that an approximate azimuthal only solution describes the flow field near the apex of the cone. They then support their abstract proof with results from simulations and experiments. In many other simulations of driven cavity flows with discontinuities, the effect of the discontinuity on the accuracy of the simulation is negligible for Reynolds numbers less than one[7], however a good explanation for this has not been provided.

Spohn et al.[2], Brons et al.[8], Mirghaie et al.[9], and Hirsa et al.[10] have studied the flow in a lid driven cavity with a free surface opposite the stationary endwall. In these studies good, agreement is only obtained between computations and experiments for aspect ratios greater than a half. In particular, Mirghaie et al.[9] show that at small aspect ratio, a surfactant on the surface of the liquid can make the flow field differ from the flow field for a surfactant free surface.

Using a regular perturbation scheme in the Reynolds number, Hills[12] has studied the first order effects of inertia in a circular semi-infinite lid driven cavity. Hills[12] found that because the zeroth order flow field is azimuthal, the first order inertial correction has a defining role in the axial and radial flow fields; it creates an infinite sequence of vortices that decay with distance from the rotating endwall.

The new results in this study are that when the dominant forcing of the secondary flow is not localized near the boundary discontinuity, the effect of the corner singular region at low Reynolds number can be neglected and the main features of the secondary flow calculated. The study also shows that as suggested by Spohn et al.[2], vortex breakdown occurs because fluid flowing up the center of the cavity passes through an adverse pressure gradient region where it can stagnate.

To obtain these results an analytical approach is used. This is because even though the solution is not valid for all Reynolds numbers, and even though it requires simplifying assumptions, when the assumptions hold, a good understanding of the dominant physical mechanisms, and an explanation for why and when a discontinuity will have a negligible effect on the flow are obtained. The results are then verified by comparison to experiment.

Following this introduction, the governing equations are obtained and scaled in section II. The flow field is solved for in Section III. The perturbation scheme used is described in section 3 part A, and exact Stokes solutions obtained in part B with a free and fixed surfaces opposite the rotating endwall. The equations governing the secondary flow are described in part C, where the effect of the discontinuity on the forcing function for the secondary flow is examined. Asymptotic solutions at small aspect ratio and small Reynolds number are obtained for the first order flow and pressure fields with a fixed endwall in parts D and E. Finally, in section IV, the results of this study are discussed in light of previous experimental and numerical work.

## II. Governing Equations

The flow of an incompressible fluid in a cylindrical lid driven cavity is modeled by the steady state Navier-Stokes equations with no slip boundary conditions which, assuming cylindrical symmetry, are in dimensionless form,

$$\text{Re}(\mathbf{u} \cdot \nabla \mathbf{u}) = \nabla^2 \mathbf{u} - \nabla P \qquad (1)$$

In this study, azimuthal symmetry is assumed. The incompressibility condition is satisfied by requiring that the velocity field is divergence free, $\nabla \cdot \mathbf{u} = 0$. With one fixed endwall, the boundary conditions are,

$$u_\theta[r,1] = r$$

$$u_\theta[r,0] = u_\theta[r,1] = u_r[r,0] = u_r[r,1] = u_z[r,0] = u_z[r,1] = 0 \qquad (2)$$

With a free surface, the boundary condition $u_\theta[r,0] = u_r[r,0] = 0$, is replaced by $\partial u_\theta/\partial z[r,0] = \partial u_r/\partial z[r,0] = 0$. As discussed in Brons et al.[8], a flat and free slip boundary condition is only appropriate for low Froude numbers for fluids with clean surfaces.

The symbols in these equations are,

$$z = z^*/L^* \qquad r = r^*/r_o^* \qquad u_\theta = u_\theta^*/\bar{u}_\theta^* \qquad P = P^*/\mu^* \bar{u}_\theta^*$$
$$\gamma = L^*/r_o^* \qquad \bar{u}_\theta^* = r_o^* \Omega^* \qquad u_r = u_r^*/\bar{u}_\theta^* \qquad u_z = u_z^*/\bar{u}_\theta^*$$
$$\text{Re} = \bar{u}_\theta^* r_o^*/\nu^*$$

The dimensional quantities have an asterisk and are: $z^*$, the axial distance from the base of the cavity, $r^*$, the radial distance from the central axis of the cavity, $r_o^*$, the radius of the cavity, $\mu^*$, the dynamic viscosity of the fluid in the cavity, $\Omega^*$, the angular velocity of the rotating lid, $\nu^*$, the kinematic viscosity of the fluid in the cavity and $P^*$, the dynamic pressure field in which the effect of a uniform gravitational field is subtracted from the real pressure field. The scaling speed, $\bar{u}_\theta^*$, is the speed of the edge of the rotating lid. The scaling lengths are the radius and length of the cavity, $r_o^*$ and $L^*$ respectively.

Although this renders the scaling near the singular region between the rotating and stationary walls invalid, for ease of computation the velocity transition between the stationary sidewall and the rotating top wall is modeled as a sudden jump

## III. Solution
### A. Perturbation expansion

An asymptotic expansion in the Reynolds number, as used by Hills[12], is used to calculate the flow. The field variables are expanded in series in powers of the Reynolds number,

$$\begin{aligned}
u_\theta &= u_{0,\theta} + \text{Re}\, u_{1,\theta} + ....\\
u_r &= u_{0,r} + \text{Re}\, u_{1,r} + ....\\
u_z &= u_{0,z} + \text{Re}\, u_{1,z} + ....\\
P &= P_0 + \text{Re}\, P_1 + ....
\end{aligned} \qquad (3)$$

Substituting these expressions into the Navier-Stokes equations and matching the equations and the boundary conditions order by order in the Reynolds number, the linear partial differential equations to be solved to determine the Stokes flow field and the first order inertial correction to the Stokes flow field are obtained.

The zeroth order Stokes flow field satisfies,

$$\begin{aligned}
\nabla^2 u_{0,\theta} &= \nabla P_0 \\
\nabla \cdot u_{0,\theta} &= 0 \\
u_{0,r} &= u_{0,z} = 0
\end{aligned} \qquad (4)$$

The no slip boundary conditions for this flow field are,

$$u_{0,\theta}[r,1] = r$$
$$u_{0,r}[1,z] = u_{0,z}[1,z] = u_{0,\theta}[r,0] = u_{0,\theta}[1,z] = 0 \qquad (5)$$
$$u_{0,r}[r,0] = u_{0,r}[r,1] = u_{0,z}[r,0] = u_{0,z}[r,1] = 0$$

With the free top surface, $u_{0,\theta}[r,0] = 0$ becomes $\partial u_{0,\theta}/\partial z[r,0] = 0$. In this Stokes equation, the azimuthal only velocity field is uncoupled from the axial and radial velocity fields.

To calculate the pressure field, pressure boundary conditions are required. Without a full solution to the nonlinear equations of motion, the appropriate boundary conditions cannot be calculated. To overcome this difficulty, which is created by using a perturbation solution, an approximation is made that all the boundaries are at atmospheric pressure. With a free surface, the pressure at the free surface is atmospheric. Furthermore, in an experiment there is a small gap between the rotating endwall and stationary sidewall. Since in steady state there is no flow through this gap, the dynamic pressure at this gap (which excludes hydrostatic effects) must be equal to the atmospheric pressure. At low Reynolds number, the dynamic pressure at the walls will not be too different from atmospheric pressure in an apparatus with a free surface. Since for aspect ratios close to one, and Reynolds numbers below 500, the secondary flows in experiments with fixed and free surfaces are the same[2], it is reasonable to assume atmospheric pressure boundary conditions as an approximation to the actual pressure boundary conditions in an experiment for both boundary conditions at low Reynolds number.

The first order correction flow field satisfies

$$u_{1,\theta} = 0$$
$$\mathbf{u}_0 \cdot \nabla \mathbf{u}_0 = \nabla^2 \mathbf{u}_1 - \nabla P_1 \qquad (6)$$
$$\nabla \cdot \mathbf{u}_1 = 0$$

With a stationary endwall, this flow field also has null boundary conditions. With a free surface, the flow field also has null boundary conditions except at the free surface where the axial partial derivative of the azimuthal and radial velocities is zero. At first order, the azimuthal velocity field is again uncoupled from the axial and radial fields. In comparison to the zeroth order flow, the first order azimuthal velocity field has no forcing term, and is zero. The first order pressure field has null boundary conditions because the approximate boundary conditions are met at zeroth order;/

A regular perturbation expansion requires that throughout the flow field, $|\mathbf{u}_0| \gg \mathrm{Re}|\mathbf{u}_1|$, $|\mathbf{u}_0| \gg \mathrm{Re}\,\mathbf{u}_0 \cdot \nabla \mathbf{u}_0$, and that $P_0 \gg \mathrm{Re}P_1$. In a regular perturbation expansion these magnitude requirements are met by scaling the velocities and distances appropriately. When this scaling cannot be accomplished, a singular perturbation solution is required and the regular perturbation solution is not uniformly valid.

**B. Stokes flow solutions**

The Stokes flow field has azimuthal symmetry and so only depends on the radial and axial coordinates, thus it automatically satisfies the continuity equation. With a fixed endwall opposite the rotating endwall, the solution, $u_{0,\theta,fix}$, is found using $rz$, and then satisfying, the no slip boundary condition on the outer stationary wall using a finite Fourier transform[13]. The final solution is,

$$u_{0,\theta,fix} = rz - 2 \sum_{n=1}^{\infty} \frac{\sin[n\pi(1-z)]}{n\pi} \frac{I_1[n\pi r/\gamma]}{I_1[n\pi/\gamma]} \tag{7}$$

where $I_1$ is the modified Bessel function of first order. Eq. (7) is plotted for three different aspect ratios in Figs. 2a, 2b, and 2c. The figures show that for tall aspect ratios, the azimuthal flow is concentrated in a region near the rotating wall, whereas for short aspect ratios the azimuthal flow has a uniform axial velocity gradient.

The series expressions for the steady state Stokes flow field are complicated, and to analytically compute the first inertial correction, a simpler expression is required. Fig. 2d shows the first part, $rz$, of the solution for an apparatus with one stationary endwall in Eq. (7). By comparing Figs. 2a and 2b to Fig. 2d, one observes that as found by Panton[4], the expression $rz$ is a good approximate representation of the Stokes flow field for small aspect ratios in an apparatus with a stationary endwall.

The simplified expression for the steady state flow field at small aspect ratio, $rz$, can also be obtained analytically. In Eq. (7) one can use the asymptotic expansions of the modified Bessel function with large argument, ($I_1[x] \sim \exp[x]/\sqrt{2\pi x}$ see Bender and Orszag[14]) to show that at small aspect ratio - which corresponds to large argument for the Bessel function – the zeroth order azimuthal velocity field, $u_{0,\theta,fix}$, is approximately equal to $rz$, everywhere except near the stationary sidewall where $r \to 1$.

To obtain the Stokes solution with a free surface, $u_{0,\theta,free}$, a correction series is added to $r$, the solution for solid body rotation, to satisfy the no slip boundary condition at the outer stationary wall. The series is found using a finite Fourier transform[13] and the full solution is,

$$u_{0,\theta,free} = r - \sum_{n=0}^{\infty} \frac{2I_1[(n+1/2)\pi r/\gamma]\sin[(n+1/2)\pi(z-1)]}{I_1[(n+1/2)\pi/\gamma](n+1/2)\pi} \tag{8}$$

This solution is plotted for aspect ratios of 0.1, 1, and 10 in Figs 3a, 3b and 3c.

With a free surface, an asymptotic analysis similar to that with a fixed end wall, shows that at small aspect ratio, the dominant solution is solid body rotation. This solution is a good approximation to the flow field at small aspect ratio and Reynolds number, everywhere except near the stationary sidewalls.

For both free and no slip boundary surfaces opposite the rotating end wall, the solution for the zeroth order dynamic pressure field is $P_0 = P_{atm}$.

**C. Governing equations for the first order inertial corrections**

The analysis in this section shows that the discontinuity between the rotating and stationary walls creates at worst a delta function forcing. Although a solution for this forcing can be found, it cannot satisfy the scaling requirements, which require that the first order flow field is smaller than the zeroth order flow field throughout the domain. Consequently, singular perturbation theory is required to resolve the flow at the junction between the rotating and stationary walls. Since it is difficult to analytically compute the flow at the junction between the rotating and stationary walls, the purpose of this section is to determine the conditions for which the effect of the discontinuity on the forcing functions can be neglected.

For both free and fixed surfaces opposite the rotating endwall, the first order azimuthal momentum equation has the solution $u_{1,\theta} = 0$. To satisfy the continuity, axial momentum and radial momentum equations, the streamfunction, $\Psi$, for the radial and axial fluid velocities is used. The streamfunction is defined such that, $u_{1,r} = (r\gamma)^{-1} \partial \Psi / \partial z$ and $u_{1,z} = -r^{-1} \partial \Psi / \partial r$. This automatically satisfies the continuity equation. Substituting these equations into the radial and axial momentum equations, and then eliminating the pressure, the following single forced fourth order linear partial differential equation is obtained,

$$2u_{0,\theta}\left(\partial u_{0,\theta} / \partial z\right) = D^2(D^2 \Psi) \tag{9}$$

where the operator $D^2$ is,

$$D^2 \equiv \frac{1}{\gamma^2} \frac{\partial^2}{\partial z^2} + r \frac{\partial}{\partial r}\left(\frac{1}{r} \frac{\partial}{\partial r}\right) \tag{10}$$

The no slip boundary conditions on the cavity walls are,

$$\frac{\partial \Psi}{\partial r}[1,z] = \frac{\partial \Psi}{\partial z}[r,0] = \frac{\partial \Psi}{\partial z}[r,1] = \frac{\partial \Psi}{\partial z}[1,z] = \frac{\partial \Psi}{\partial r}[r,0] = \frac{\partial \Psi}{\partial r}[r,1] = 0 \tag{11}$$

The forcing function with one stationary endwall is,

$$2u_{0,\theta,fix}\frac{\partial u_{0,\theta,fix}}{\partial z}$$

$$= 2\left(rz - 2\sum_{n=1}^{\infty}\frac{\sin[n\pi(1-z)]}{n\pi}\frac{I_1[n\pi r/\gamma]}{I_1[n\pi/\gamma]}\right)\left(r + 2\sum_{n=1}^{\infty}\cos[n\pi(1-z)]\frac{I_1[n\pi r/\gamma]}{I_1[n\pi/\gamma]}\right) \quad (12)$$

This forcing function does not converge in the classical sense. The convergence properties are worst at the discontinuity between the rotating and stationary walls. By examining the properties of the forcing function at the discontinuity, one can check whether a reasonable solution to the first order flow can be obtained. At the rotating outer wall, the most singular part of the forcing function becomes the product of a Dirac delta function and a Heaviside function. Products of distributions with support at the same point are ill defined; however the magnitude of the Heaviside function is bounded, and so the strength of the forcing is at worst a delta function forcing,

To obtain a physically reasonable solution for the first order flow, the conditions at the corner cannot be modeled using a jump discontinuity. Firstly, because the modeling assumption that the first order flow has a smaller magnitude than the zeroth order flow is violated. Secondly, because in any physical situation, there will be an imposed length scale appropriate to the corner, this could be: a balance between viscous and inertial forces, a balance between surface tension and viscous forces, the gap between the rotating and stationary walls, or a Knudsen length scale because the continuum assumption no longer holds. It is therefore appropriate to modify the forcing function to reflect a minimum resolvable scale.

An approximate forcing function can be obtained by truncating the series representation. This imposes a minimum length scale. Simple truncation does not ensure a convergent series. In a trigonometric expansion, a procedure that results in a convergent truncated series is the application of Lanczos correction factors to the original trigonometric expansion[15]. These factors replace the discontinuous jump in velocity, with a smooth rapid transition region that still preserves the properties of the solution away from the discontinuity. The factors also preserve the integral value of the forcing. The forcing function is then given by,

$$2u_{0,\theta,fix}\frac{\partial u_{0,\theta,fix}}{\partial z} = 2\left(rz - 2\sum_{n=1}^{k}\frac{\sin[n\pi(1-z)]}{n\pi}\frac{\sin(n\pi/k)}{n\pi/k}\frac{I_1[n\pi r/\gamma]}{I_1[n\pi/\gamma]}\right)$$

$$\times\left(r + 2\sum_{n=1}^{k}\cos[n\pi(1-z)]\frac{\sin(n\pi/k)}{n\pi/k}\frac{I_1[n\pi r/\gamma]}{I_1[n\pi/\gamma]}\right) \quad (13)$$

The index $k$ denotes the largest resolvable wave number. The forcing function is now a convergent sum over a finite number of terms. Forcing functions for aspect ratios of 0.1, 1, and 10 are shown in Figs. 4a, 4b, and 4c. From these figures one observes that at the transition between the rotating and stationary walls, the forcing function has a large positive value which, when a large number of terms are used in the series representation, approaches a delta function.

Fig. 4d shows the approximate forcing function for a flow with a stationary endwall, $2r^2z$, with the most singular part removed. It is quantitatively similar to the forcing function at an aspect ratio of 0.1 (Fig. 4a) and qualitatively similar to the forcing function at an aspect ratio of one (Fig. 4c). The approximate forcing function differs from the exact forcing functions at the boundary discontinuity, where the exact forcing functions have a large magnitude spike.

With a free surface, the series expansion for the forcing function also does not converge because of the discontinuity between the rotating endwall and the stationary sidewall. Using Lanczos correction factors, the forcing function is,

$$u_{0,\theta,free}\frac{\partial u_{0\theta,free}}{\partial z} = \left(r - \sum_{n=0}^{k}\frac{2I_1[(n+1/2)\pi r/\gamma]}{I_1[(n+1/2)\pi/\gamma]}\frac{\sin[(n+1/2)\pi(z-1)]}{(n+1/2)\pi}\frac{\sin((n+1/2)\pi/k)}{(n+1/2)\pi/k}\right)$$
$$\times\left(-\sum_{n=0}^{k}\frac{2I_1[(n+1/2)\pi r/\gamma]}{I_1[(n+1/2)\pi/\gamma]}\frac{\sin((n+1/2)\pi/k)}{(n+1/2)\pi/k}\cos[(n+1/2)\pi(z-1)]\right) \quad (14)$$

This forcing function is plotted in Figs. 5a, 5b and 5c for aspect ratios of 0.1, 1 and 10. At small aspect ratio, Fig. 5a shows that the forcing function is close to zero everywhere apart from near the stationary sidewall. This behavior is very different from the forcing function shown in Fig. 4a for an apparatus with a stationary endwall opposite the rotating endwall; in which the forcing is not localized near the junction between the rotating and stationary walls. Figs. 5b and 5c however, are similar to Figs. 4b and 4d for which the free surface is replaced by a no slip boundary condition.

**D. The first order flow field at small aspect ratio with a fixed endwall**

The first order flow field is computed using the approximate small aspect ratio Stokes solution for an apparatus with a stationary endwall, $rz$, as the forcing function. The first order flow field is not computed for a small aspect ratio apparatus with a free surface because in this geometry the flow near the boundary discontinuity must be computed to calculate the internal axial and radial motions. The partial differential equation for the particular solution of the streamfunction is, $2r^2z = D^2(D^2\Psi^p)$. The appropriate boundary conditions are discussed in part C, however these are modified to obtain an analytically tractable solution; a free slip boundary condition is enforced on the top and bottom walls for the radial motions. $\Psi^p[r,1] = \Psi^p[r,0] = 0$. This boundary condition does not hold in an experiment, but because the driving force for this flow does not occur at the walls, the solution obtained gives the qualitative features of the secondary flow pattern.

A solution for the secondary flow field is found using a finite Fourier transform[13] with sinusoidal eigenfunctions in the axial direction. The homogeneous solutions in the radial direction that are finite at the origin, $(r/\gamma)I_1[n\pi r/\gamma]$ and $(r^2/\gamma^2)I_0[n\pi r/\gamma]$, are added to the required particular solution of the resulting differential equation, $8r^2/(n\pi)^5$. The full result is,

$$\Psi[r,z] = \sum_{n=1}^{\infty} \frac{2\sin[n\pi z]}{n\pi}\left(G1_n \frac{r}{\gamma}I_1[n\pi r/\gamma] + G2_n \frac{r^2}{\gamma^2}I_0[n\pi r/\gamma] + 4\frac{r^2}{(n\pi)^4}\right) \quad (15)$$

The constants $G1_n$ and $G2_n$, are chosen to ensure that the streamfunction and its first radial partial derivative are equal to zero on the outer stationary wall.

Sample streamfunctions are shown in Figs. 6a and 6b for aspect ratios of 0.1 and 1. The centrifugal forces created by the rotating top wall push fluid out away from it. The fluid then travels down in a narrow layer along the stationary outer wall before returning to the top rotating layer through the center of the cavity.

The calculated streamline patterns in Figs. 6a and 6b are similar to those observed in experiments for Reynolds numbers below a thousand[1]. Experiments show that for aspect ratios less than one, and for Reynolds numbers less than a thousand, vortex breakdown does not occur[1,2]. These experiments also show that the same flow pattern exists for Reynolds numbers less than one and up to a thousand before vortex breakdown[1,2]. This justifies the approximations made in obtaining the analytical solution.

**E. The first order pressure field at small aspect ratio with a fixed endwall**

The first order pressure field is governed by an equation obtained by taking the divergence of the first order equations of motion. This gives,

$$\frac{2u_{0,\theta}}{r}\frac{\partial u_{0,\theta}}{\partial r} = \frac{1}{r}\frac{\partial}{\partial r}\left(r\frac{\partial P_1}{\partial r}\right) + \frac{1}{\gamma^2}\frac{\partial^2 P_1}{\partial z^2} \quad (16)$$

As described in section III part A, null boundary conditions are imposed on all walls for this forced, second order, linear partial differential equation. Like the forcing function for the flow field, the series representation of the forcing function for the pressure field also does not converge because of the discontinuous boundary conditions. As was shown in section III part B, $rz$ is an asymptotic representation of the velocity field in a small aspect ratio apparatus with a stationary endwall. When this is used in the forcing function, the effect of the boundary discontinuity is neglected and a tractable analytical solutions is possible. The resulting partial differential equation for the pressure field is,

$$2z^2 = \frac{1}{r}\frac{\partial}{\partial r}\left(r\frac{\partial P_1}{\partial r}\right) + \frac{1}{\gamma^2}\frac{\partial^2 P_1}{\partial z^2} \quad (17)$$

A solution to this equation is obtained using a finite Fourier-Bessel transform[13] in the radial direction. The resulting ordinary differential equation has the homogeneous solutions, $\sinh[\lambda_n \gamma z]$ and $\cosh[\lambda_n \gamma z]$, and a particular solution $-2\sqrt{2}J_1[\lambda_n]\lambda_n^{-3}(z^2 + 2\gamma^{-2}\lambda_n^{-2})$. The full solution is,

$$P_1 = \sum_{n=1}^{\infty} 4\lambda_n^{-3} J_0[\lambda_n r] \left( \frac{\sinh[\lambda_n \gamma z]}{\sinh[\lambda_n \gamma]} \left(1 + \frac{2(1-\cosh[\lambda_n \gamma])}{\gamma^2 \lambda_n^2}\right) + 2\gamma^{-2}\lambda_n^{-2}\cosh[\lambda_n \gamma z] - z^2 - 2\gamma^{-2}\lambda_n^{-2} \right) \qquad (18)$$

This solution is plotted in Figs. 7a and 7b for aspect ratios of $\gamma = 0.1$ and $\gamma = 1$.

## IV. Discussion
### A. Comparison with Semi-infinite Cavity

Hills[12] showed that for a semi-infinite cylindrical lid driven cavity, the primary azimuthal flow field decayed exponentially with distance from the rotating endwall. This is in agreement with the results from this study where in a large aspect ratio apparatus, the primary azimuthal flow is concentrated near the rotating endwall.

Also in agreement with this study, Hills[12] found that the centrifugal forces created by the rotating endwall drive the axial and radial fluid motions. For the semi-infinite cavity however, Hills[12] found a sequence of eddies that decayed with distance from the endwall. This study shows that in a small aspect ratio apparatus with a stationary endwall, there is only one eddy in the secondary flow. Comparing these results to studies of the rectangular lid driven cavity[7], suggests that as the aspect ratio of the cylindrical lid driven cavity is increased, the number of eddies will also increase.

This study shows that as found by Hills[12], because the primary flow field has only an azimuthal component, the first order correction to the primary flow field has a determining influence on the axial and radial velocity fields. Similarly, because the gradient of the primary dynamic pressure field is zero, the first order correction to the pressure field has a determining influence on the pressure gradient.

### B. Why vortex breakdown occurs

The flow map obtained by Escudier[1] shows that with one fixed endwall, vortex breakdown only occurs for aspect ratios greater than one. This calculation shows that the vortex is created by centrifugal forcing due to the rotating endwall. This calculation supports Spohn et al.'s[2] suggestion that the vortex breaks down because of the adverse pressure gradient faced by the secondary fluid flow in the center of the cavity. This is because an adverse pressure gradient exists even at very small Reynolds numbers and the flow pattern at small Reynolds numbers is similar to that at Reynolds numbers just below those for vortex breakdown.

At very small aspect ratio, Figs. 7a and 7b show that even though there is an adverse pressure gradient, the adverse pressure gradient is very small, and so will not be large enough to cause vortex breakdown in an apparatus with a stationary endwall opposite the rotating endwall. The streamlines in Figs. 6a and 6b also show that the secondary flow along the central axis of the cavity gets weaker with increasing aspect ratio because the streamlines are wider. This is because with increasing aspect ratio, the dissipative flow is driven the same amount by the rotating endwall, but needs to travel a longer distance to circulate round the cavity.

Vortex breakdown has also been observed in lid driven cavities with a free surface for aspect ratios greater than a half[2]. The forcing functions for axial and radial flow for lid driven cavities with aspect ratios greater than one in and with free slip or no slip boundaries opposite the rotating endwall are similar; this suggests that the vortex breakdown mechanism is the same for the two geometries. For very small aspect ratios however, the forcing functions are different, suggesting that the secondary flow in a small aspect ratio apparatus with a free surface - even if it is clean - will be very different from the simple circulation pattern calculated here for an apparatus with a fixed endwall.

**C. Boundary Discontinuity**

The boundary conditions examined in this study have a discontinuity at the junction between the rotating and stationary walls where it is difficult to compute the flow field. In many numerical studies of lid driven cavities, the computed flow field is also insensitive to the modeling of the discontinuity[7]. The results of this study show that this insensitivity is because the distributed forcing by the moving lid, dominates over the forcing by the corner discontinuity. In such a situation, the effect of the discontinuity on the secondary flow can be neglected entirely.

In contrast, when the forcing is localized near the discontinuity, such as in the small aspect ratio apparatus with a free slip surface, the discontinuity can be important and this may explain why simulations fail to reproduce experimental results[8]. Furthermore, because the forcing by the discontinuity is highly localized, any other disturbances can have a large impact on the flow field. In particular, because in a small aspect ratio apparatus with a free surface a significant portion of the fluid is in solid body rotation and has weak axial and radial currents, surface stresses due to surfactants can have an important influence on the flow field. This may explain why when inertial effects are important, the flow field with a surfactant is very different from the flow field without a surfactant as observed by Mirghaie et al.[9].

Once the eddy structure observed at low Reynolds number ceases to exist, Sinhas'[3] experimental study shows that the creation of extra vorticity at the discontinuity between the rotating and stationary endwalls changes the flow pattern in the bubble. When the bubble is formed, fluid that has flowed past the discontinuity stagnates in the middle of the cavity. The vorticity created by the discontinuity changes the flow field in the bubble because it affects the vorticity distribution at the stagnation point. Since there is no axial motion at the stagnation point, the vorticity at the stagnation point has an important influence on the flow field in the bubble.

Spohn et al.[2] compared their experimental observations of the flow field in a cavity with a fixed endwall to numerical simulations. They observed that after vortex breakdown, simulations produced radially symmetric bubbles whereas experiments produced radially asymmetric bubbles. This might be due to the effect of local asymmetries at the discontinuity between the rotating endwall and stationary sidewall.

Finally, if one attempts to obtain a regular perturbation solution in the Reynolds number for a circular Couette flow in a finite apparatus with stationary endwalls: the Stokes flow solution can be found using a finite Fourier transform[16,17]; the dominant component of the forcing function for the first inertial correction however, is at the corner discontinuity. In such a situation one cannot neglect the discontinuity. Experiments and simulations of circular Couette flows with a rotating inner cylinder, a stationary outer cylinder, and stationary endwalls, have anomalous modes where the secondary flow is axially asymmetric, even though the domain is axially symmetric[18]. When experiments are done with a rotating inner cylinder, stationary outer cylinder, and rotating endwalls, anomalous modes are not observed[18]. This is because with a rotating endwall, the centrifugal forces dominate in driving the secondary flow, whereas with a stationary endwall, the flow field near the discontinuity drives the secondary flow. In such a situation, local asymmetries at the discontinuity may give rise to these anomalous symmetry breaking modes.

## V. Conclusion

The analytical solution found in this study shows that for Reynolds numbers less than one, when the axial and radial motions are primarily driven by distributed centrifugal forces created by the rotating endwall, the flow in a lid driven cavity can be computed without a detailed description of the conditions at the boundary discontinuity. When inertial effects are important, the conditions at the discontinuity may have a strong effect on the flow field and need to be carefully modeled for simulations to replicate experimental observations. The results also support Spohn et al.'s[2] suggestion that, vortex breakdown in this geometry is due to stagnation of the secondary flow by an adverse pressure gradient.

## Acknowledgements

The author would like to thank Professor Alexander Smits for some very helpful comments on an earlier draft of this paper. A portion of this study was completed as part of the requirements of a master of science in engineering degree at Princeton University, during which the author was supported by an Upton Fellowship.

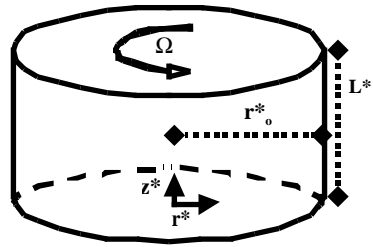

**Fig. 1. A circular lid driven cavity.**

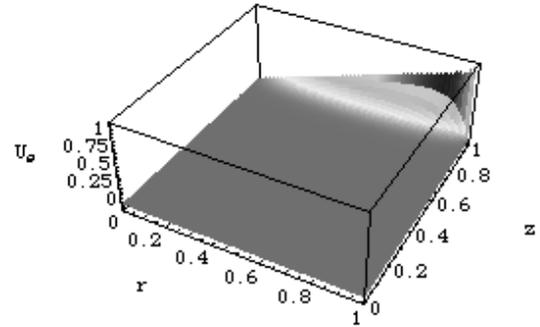

**Fig. 2c. Stokes solution for an aspect ratio γ = 10 with a stationary endwall.**

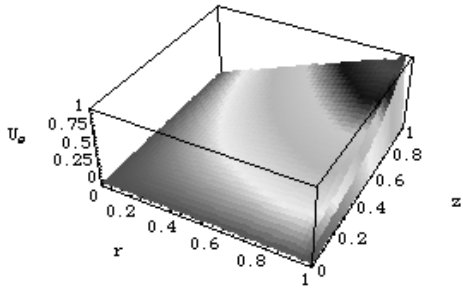

**Fig. 2a. Stokes solution for an aspect ratio γ = 0.1 with a stationary endwall.**

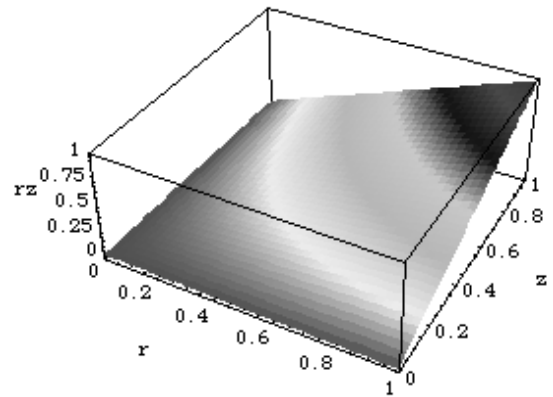

**Fig. 2d. A plot of rz.**

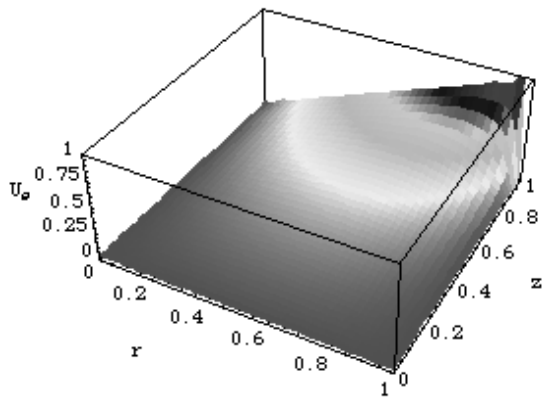

**Fig. 2b. Stokes solution for an aspect ratio γ = 1 with a stationary endwall.**

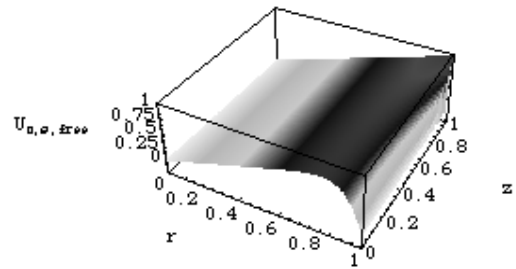

**Fig. 3a. Stokes solution for an aspect ratio γ = 0.1 with a free surface.**

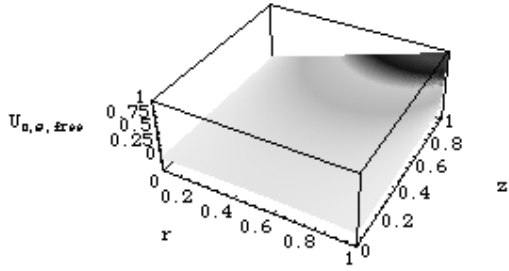

**Fig. 3b. Stokes solution for an aspect ratio γ = 1 with a free surface.**

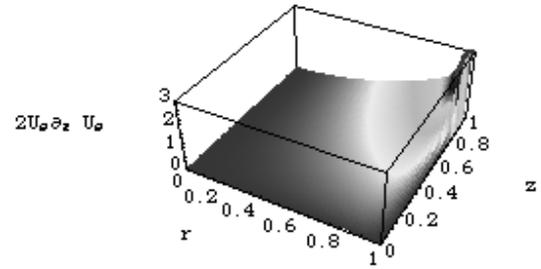

**Fig. 4a. The forcing function for the secondary flow for an aspect ratio γ = 0.1 with a stationary endwall.**

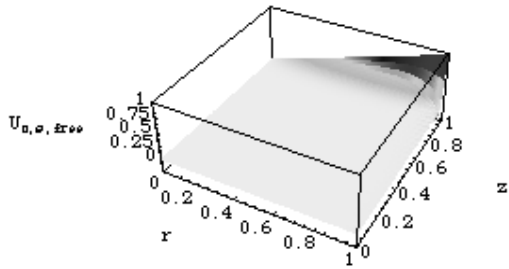

**Fig. 3c. Stokes solution for an aspect ratio γ = 10 with a free surface.**

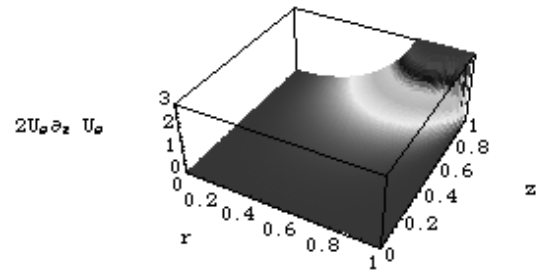

**Fig. 4b. The forcing function for the secondary flow for an aspect ratio γ = 1 with a stationary endwall.**

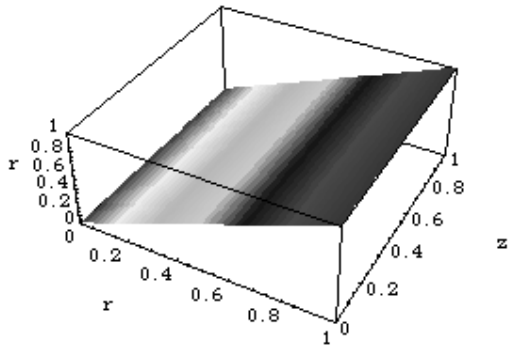

**Fig. 3d. A plot of r.**

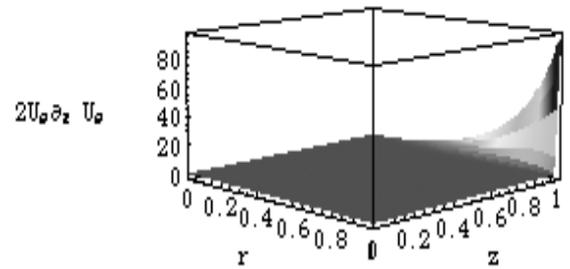

**Fig. 4c. The forcing function for the secondary flow for an aspect ratio γ = 10 with a stationary endwall.**

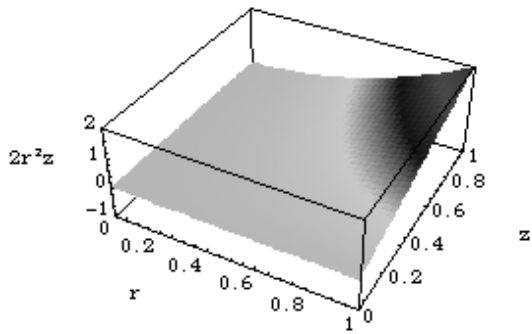

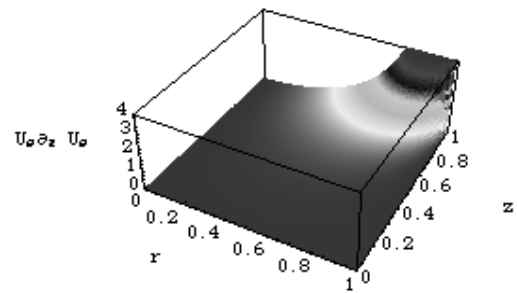

**Fig. 4d.** The approximate forcing function for the secondary flow when the most singular component is removed.

**Fig. 5b.** The forcing function for the secondary flow for an aspect ratio $\gamma = 1$ with a free surface.

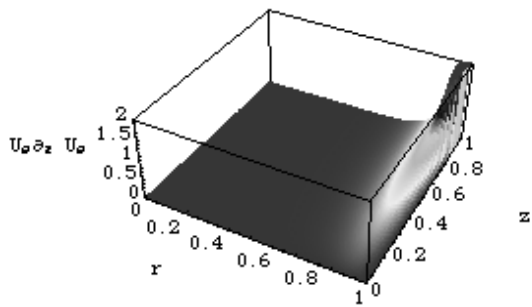

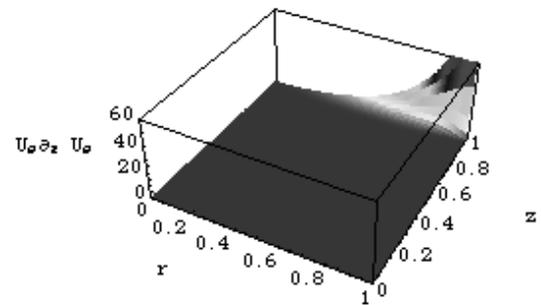

**Fig. 5a.** The forcing function for the secondary flow for an aspect ratio $\gamma = 0.1$ with a free surface.

**Fig. 5c.** The forcing function for the secondary flow for an aspect ratio $\gamma = 10$ with a free surface.

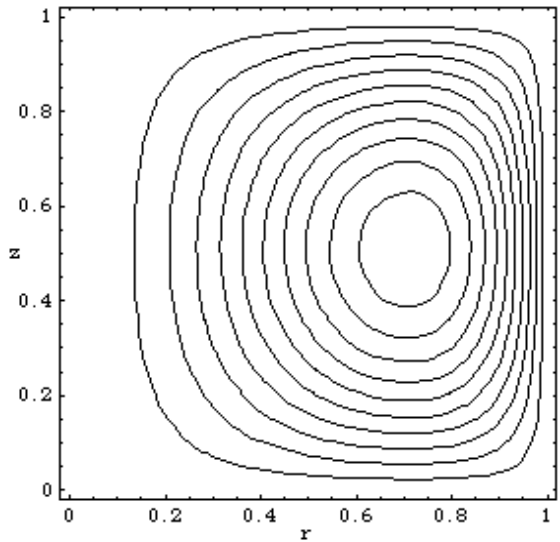

**Fig. 6a. The streamfunction for axial and radial fluid motions at an aspect ratio $\gamma = 0.1$.**

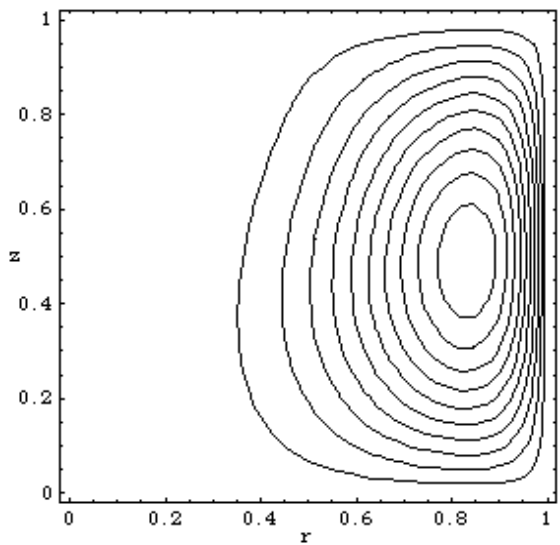

**Fig. 6b. The streamfunction for axial and radial fluid motions at an aspect ratio $\gamma = 1$.**

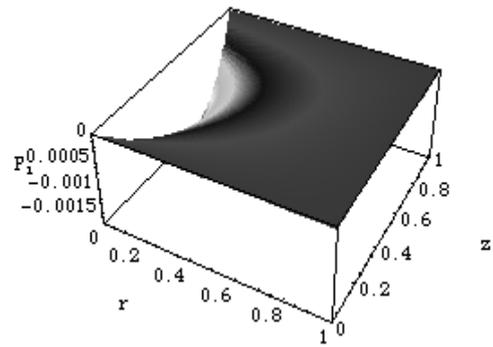

**Fig. 7a. The first order correction to the pressure field in the circular lid driven cavity at an aspect ratio $\gamma = 0.1$.**

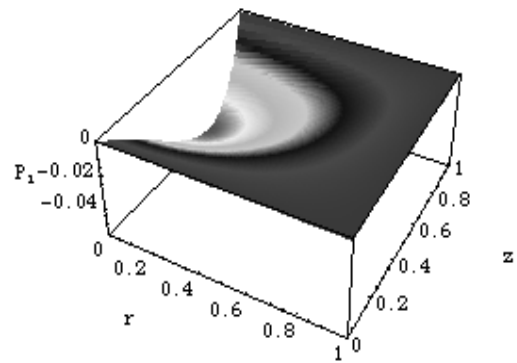

**Fig. 7b. The first order correction to the pressure field in the circular lid driven cavity at an aspect ratio $\gamma = 1$.**